\documentclass[aps,prl,preprint,superscriptaddress]{revtex4-1}
\usepackage{amsmath} 
\usepackage{graphicx}
\usepackage{hyperref}
\newcommand{\dint}{\mathrm{d}}

\newcommand{\be}{\begin{equation}} \newcommand{\ee}{\end{equation}}

\newcommand{\ben}{\begin{enumerate}} \newcommand{\een}{\end{enumerate}}
\newcommand{\bc}{\begin{center}} \newcommand{\ec}{\end{center}}
\newcommand{\bi}{\begin{itemize}} \newcommand{\ei}{\end{itemize}}

\begin{document}


\title{Greedy algorithms and Zipf laws}


\author{Jos\'{e} Moran}
\affiliation{Centre d'Analyse et de Math\'{e}matiques Sociales, EHESS, 54 Boulevard Raspail, 75006 Paris}
\email{jose.moran@polytechnique.org}
\author{Jean-Philippe Bouchaud}
\affiliation{Capital Fund Management, 23 Rue de l'Universit\'{e}, 75007 Paris}


\date{\today}

\begin{abstract}
We consider a simple model of firm/city/etc. growth based on a multi-item criterion: whenever entity B fares better that entity A on a subset of $M$ items out of $K$, the agent originally in A moves to B. We solve the model analytically in the cases $K=1$ and $K \to \infty$. The resulting stationary distribution of sizes is generically a Zipf-law provided $M > K/2$. When $M \leq K/2$, no selection occurs and the size distribution remains thin-tailed. In the special case $M=K$, one needs to regularise the problem by introducing a small ``default'' probability $\phi$. We find that the stationary distribution has a power-law tail that becomes a Zipf-law when $\phi \to 0$. The approach to the stationary state can also been characterized, with strong similarities with a simple ``aging'' model considered by Barrat \& M\'ezard. 
\end{abstract}


\maketitle

\section{Introduction}
\label{sec:description}
Among all power-law distributions, the Zipf law appears to play a special role. Mathematically, the Zipf law corresponds to a marginally (logarithmically) divergent first moment -- distributions decaying faster have a finite mean, while distributions decaying slower have an infinite mean. Empirically, this law appears to hold in a variety of situations, from the frequency of words to the node degree of the Internet~\cite{zipfinternet,huberman99evolutionary}, the size of companies ~\cite{Axtell2001} or the population of cities~\cite{gabaixcities,saichev2009theory}; for reviews see~\cite{gabaix2009power,saichev2009theory}. Correspondingly, there has been a flurry of possible explanation for the ubiquitous Zipf law. One standard explanation for the Zipf law is based on stochastic multiplicative growth, which under relatively mild assumptions (weak redistribution between entities and/or total mass conservation) can be shown to converge towards a Zipf distribution~\cite{gabaixcities,bouchaudmezard2000,saichev2009theory}. More recently, Gualdi \& Mandel~\cite{gualdi2016emergence} and Axtell~\cite{axtell2013endogenous} have proposed a different mechanism in the context of company sizes, based on the idea that larger firms are more ``attractive'' than smaller firms, and individual agents tend to switch from small firms to large firms. 

Inspired by these ideas, we propose here a schematic model where agents randomly select over time better and better items, possibly using multiple criteria. For definiteness, we will think about firms, but other interpretations are
possible. We postulate that each firm $i$ is characterized by a set of $K$ attributes, with scores $\mathbf{x}^i \in [0,1]^K$ that all agents agree on. For instance, $x^i_1$ is the score given to the wages paid by firm $i$, $x^i_2$ the score given to its location, $x^i_3$ the score given to its work environment, etc. Firm $j$ is deemed ``better'' than firm $i$ when the number of attributes for which $j$ has a higher score than $i$ exceeds $M$ (with $1\leq M\leq K$).
In the single criterion case $K=1$, we find that the stationary distribution of firm sizes is a Zipf distribution when the reset rate tends to zero. Other power-laws are possible, depending on the specific search mechanism. In the multi-criterion case, we show that the asymptotic distribution is a truncated Zipf law as soon as $M > K/2$ (up to logarithmic corrections when $K > 1$), with a diverging cutoff in the limit $K \to \infty$, $M/K > 1/2$. For $M \leq K/2$, there is no ``condensation'' and hence no power-law tails, as expected.

\section{The single criterion model}

\subsection{Greedy algorithm} 

We first consider the case $K=1$. Each firm $i=1,\dots, N$ is characterized by a single score $x^i \in [0,1]$. Firms are populated with agents; the fraction of agents belonging to firm $i$ at time $t$ is denoted $n_i(t)$. At each time infinitesimal step ${\rm d}t$, an agent is chosen at random with probability $\Gamma {\rm d}t$; s/he picks at random a firm among the ones that are ``better''  than the firm s/he presently works for, and moves to the newly chosen firm. We consider from the outset the limit $N \to \infty$ and assume without loss of generality that the $x^i$ are distributed uniformly in $[0,1]$. (One can always transform any chosen distribution of $x$ into the uniform one by a change of variable that preserves the ordering of the scores.) We also choose to label firms by their score $x^i$ rather than by $i$, and describe the system at time $t$ by the function $n(x,t)$ such that the fraction of agents belonging to firms with score between $x$ and $x + {\rm d}x$ is 
$n(x,t) {\rm d}x$. The evolution of $n(x,t)$ when agents use the greedy algorithm above is, for $x < 1$:
\be\label{eq:master1}
\frac{\partial n(x,t)}{\partial t} = - \Gamma n(x,t) + \Gamma \int_0^x {\rm d}y \frac{n(y,t)}{1-y},
\ee
where the second term in the RHS corresponds to agents in firms with score $y < x$ choosing their next firm in the interval $[y,1]$ with uniform probability. As written, Eq. \eqref{eq:master1} obviously leads to a stationary state where all agents condense in the firm with the highest score (i.e. $x=1$ for $N \to \infty$.) Some regularization is needed to make the problem non-trivial and interesting. One realistic assumption is that each firm has a probability $\varphi$ per unit time to go under, in which case all its employees find new jobs with uniform probability in the remaining firms. The defaulted firm is replaced by a new one with an $x$ chosen uniformly in $[0,1]$. This leads to the following modified Master equation
\be\label{eq:master1bis}
\frac{\partial n(x,t)}{\partial t} = - \Gamma n(x,t) + \Gamma \int_0^x {\rm d}y \frac{n(y,t)}{1-y} - \varphi n(x,t) + \varphi.
\ee
Another, nearly equivalent interpretation, is to assume that new agents are injected in a growing economy, with a constant rate $\varphi$ and a uniform ``deposition rate'' among all firms. 

We first look for a stationary solution to Eq. \eqref{eq:master1bis}. Setting the LHS to zero and changing variable to $u=1-x$, the stationary solution $n_{\text{st}}(u)$ must obey the following differential equation:
\be
-\Gamma n_{\text{st}}^\prime(u) - \Gamma \frac{n_{\text{st}}(u)}{u} - \varphi n_{\text{st}}^\prime(u) = 0.
\ee
Introducing $\phi:=\varphi/\Gamma$, the normalized solution to the above equation reads:
\be\label{eq:greedy_stat_sol}
n_{\text{st}}(u) = \frac{\phi}{1+\phi}\frac{1}{u^{\frac{1}{1+\phi}}}.
\ee
This equation gives us the equilibrium fraction of agents in firms of quality $x=1-u$.

We are now in position to derive the stationary distribution of the size $s$ of firms. Since the distribution of $u$'s is uniform, one has
\be
P(s) = \int_0^1 {\rm d}u \, \delta\left(s - \frac{\phi}{1+\phi}\frac{1}{u^{\frac{1}{1+\phi}}}\right) = \phi \left(\frac{\phi}{1+\phi}\right)^{1+\phi} s^{-2-\phi},\quad (s > \phi/(1+\phi))
\ee
which tends to a Zipf law when $\phi \to 0$. For companies, Axtell's results translate to $\phi \approx 0.06$. If one takes $\varphi=0.07$/year as in \cite{daepp20150120}, this suggests a reasonable search frequency of $\Gamma \approx 1$/year, although empirical data suggests a median job tenure of about $4$ years in the United States \cite{jobtenure}. 

Note that our model is a continuous formulation of the SSR process studied by Corominas-Murtra et al. in ~\cite{corominas2017sample}. It is also closely related to the well known Simon model for growing networks ~\cite{simon1955class}, where each new site links proportionally to the size of pre-existing clusters with probability $1/(1+\phi)$ and creates a new singleton cluster with probability $\phi/(1+\phi)$.

One can in fact obtain the full time dependent solution of Eq. \eqref{eq:master1bis} as follows: we first write $n(u,t) = n_{\text{st}}(u) + m(u,t) e^{-\varphi t}$ with $\int \dint u m(u,t)=0$. Substituting in Eq. (\ref{eq:master1bis}) one finds that $m(u,t)$ obeys exactly the un-regularized equation \eqref{eq:master1}. The physical interpretation of this equation is very simple: it describes the multiplicative process where the random 
variable $u(t+ {\rm d}t)$ is equal to $u_t$ with probability $1 - \Gamma {\rm d}t$, and to $a_t u_t$ with probability $\Gamma {\rm d}t$, where $a_t$ is a random variable with uniform distribution in $[0,1]$. Hence, the quantity
$\ell := -\log(u)$ is an additive random walk, for which the probability distribution evolves according to a standard Fokker-Planck equation:
\begin{equation}\label{eq:fp_eq}
 \begin{split}
 \frac{\partial P(\ell,t)}{\partial t}&= - \Gamma \frac{\partial P(\ell,t)}{\partial \ell} + \frac{\Gamma}{2} \frac{\partial^2 P(\ell,t)}{\partial \ell^2}
 \end{split}
 \end{equation}
Hence:
\be
m(u,t) = \frac{1}{u} \int {\rm d}\xi \frac{M_0(\xi)}{\sqrt{2\pi \Gamma t}} \exp\left(-\frac{(\log u + \xi + \Gamma t)^2}{2 \Gamma t}\right),
\ee
where $M_0(\xi)$ is the initial condition, such that $u m(u,0)=M_0(-\log u)$. 

\subsection{Greedy but Myopic}
\label{ss:greedy_myopic}
In the model above, we assume that each agent preselects ``better'' firms and makes a move with probability 1 if chosen. Assume now that if chosen our agent picks a firm at random among all possible firms, and decides to move only if the chosen firm is better than the current one. An agent in a firm of quality $x=1-u$ will thus only move with probability $u$. The corresponding Master equation now reads:
\be\label{eq:master2bis}
\frac{\partial n(x,t)}{\partial t} = - \Gamma n(x,t)(1-x) + \Gamma \int_0^x {\rm d}y \, n(y,t) - \varphi n(x,t) + \varphi.
\ee
Its stationary solution now reads:
\be
n_{\text{st}}(u) = \frac{\phi(\phi+1)}{(u+\phi)^2},
\ee
and the stationary distribution of firm sizes is:
\be
P(s) = \int_0^1 {\rm d}u \, \delta\left(s - \frac{\phi(\phi+1)}{(u+\phi)^2}\right) = \frac{\sqrt{\phi(1+\phi)}}{2} s^{-3/2}
\ee
where the last equality holds in the range $s \in [\phi/(1+\phi),1 + 1/\phi]$, outside which $P(s)$ is zero. Hence we find that in this model, the tail of the distribution is {\it fatter} than that of the Zipf law. This (perhaps counter-intuitive) result is due to the fact that the time spent by each agent in large firms is longer in the present setting, due to the myopic search algorithm that becomes extremely inefficient when $u \to 0$.\footnote{The reader may realize that the difference between the two models is precisely related to the difference between naive Monte-Carlo and the so-called Bortz-Kalos-Lebowitz algorithm.}

The full dynamics of the model can also be solved. Using Laplace transforms and the method of characteristics, the deviation from equilibrium is given by:
\begin{equation}\label{eq:delta_solution}
m(u,t)e^{-\varphi t}=e^{-(u+\phi)\Gamma t}m(u,0)+ \Gamma t e^{-(u+\phi) \Gamma t}\int_{u}^{1}\mathrm{d}v\, m(v,0),
\end{equation}
indicating that the system converges to the stationary state $n_{\text{st}}(u)$ after a time $\sim (u \Gamma)^{-1}$ for $u \gg \phi^{-1}$ and after a time $\sim \varphi^{-1}$ for small $u$'s. Hence, $\phi^{-1}$ acts as both a cut-off limiting the values attainable by $s$, and as the relaxation time-scale of the system. 

When $\varphi=0$, Eq. \eqref{eq:master2bis} is in fact equivalent to a model studied by Barrat and M\'{e}zard in ~\cite{barrat1995phase}, where a particle moves in a landscape of randomly distributed energy levels. At zero temperature, the energy can only go down and the Master equation of the Barrat and M\'{e}zard model reads
\begin{equation}\label{eq:BM}
\begin{split}
  \partial_t p(E,t)&=-p(E,t)\int_{-\infty}^{E}\dint E'\,\mathcal{P}(E')+\int_{E}^{\infty}\dint E'\,\mathcal{P}(E')p(E',t)
  \end{split}
\end{equation}
where $p(E,t)$ is the probability to find the system at energy $E$ at time $t$ and $\mathcal{P}(E)$ is the density of energy states. Eq. (\ref{eq:BM}) is equivalent to our model through the mapping $u=\int_{-\infty}^{E}\dint E'\,\mathcal{P}(E')$. In terms of $u$, the exact dynamical solution is given by:
\begin{equation}\label{eq:nonreg_solution}
\begin{split}
n(u,t)=e^{-\Gamma ut}n(u,0)+\Gamma t e^{-u \Gamma t}\int_{u}^{1}\mathrm{d}u\, n(u,0).
\end{split}
\end{equation}
This solution becomes universal at large times in the regime where $u = z/\Gamma t$, $z$ finite. One finds:
\be
n(u,t) = \Gamma t \mathcal{F}(u \Gamma t); \qquad \mathcal{F}(z)= e^{-z}.
\ee
We also compute the distribution $p_t(\tau)$ of trapping times $\tau:=\frac{1}{\Gamma  u}$ at time t, given by:
\begin{equation}\label{eq:trapping_times}
   p_t(\tau)=\frac{t}{\tau^2}e^{-\frac{t}{\tau}}
\end{equation}
in full agreement with the findings of ~\cite{barrat1995phase} in the appropriate limit $t \to \infty$, $\tau \to \infty$ with $t/\tau = O(1)$. In our context, the trapping time represents the time spent by an agent in a firm before moving to another firm. 
\subsection{Interpolating between models}

The two models above can be cast in a single framework, where agents spend a $u$-dependent time $\tau(u)$ in their firm before looking for a better firm. The general Master equation then reads
\be\label{eq:master3bis}
\frac{\partial n(u,t)}{\partial t} = - \frac{n(u,t)}{\tau(u)} + \int_u^1 {\rm d}v \, \frac{n(v,t)}{v \tau(v)} - \varphi n(u,t) + \varphi.
\ee
The case $\tau(u)=\Gamma^{-1}$ corresponds to the Greedy model introduced previously, while the Greedy but myopic model is when $\tau(u)=(\Gamma u)^{-1}$. One can interpolate between the two models by choosing 
$\tau(u)=\Gamma^{-1} u^{-\beta}$. The stationary solution is then found to be, for $\beta\neq0$,
\be
n_{\text{st}}(u) = \frac{Z_\beta(\phi)}{(u^\beta+\phi)^{1+\frac{1}{\beta}}},
\ee
where $Z_\beta$ is a normalisation, and the solution given in Eq. \eqref{eq:greedy_stat_sol} for $\beta=0$. The corresponding tail of firm size distribution behaves as $P(s) \sim s^{-\frac{\beta+2}{\beta+1}} $, recovering the Zipf law for $\beta=0$ and the $s^{-3/2}$ behaviour for $\beta=1$. Note that the above result continues to make sense as long as $\beta > -1$.

\begin{figure}
\includegraphics[width=1.0\textwidth]{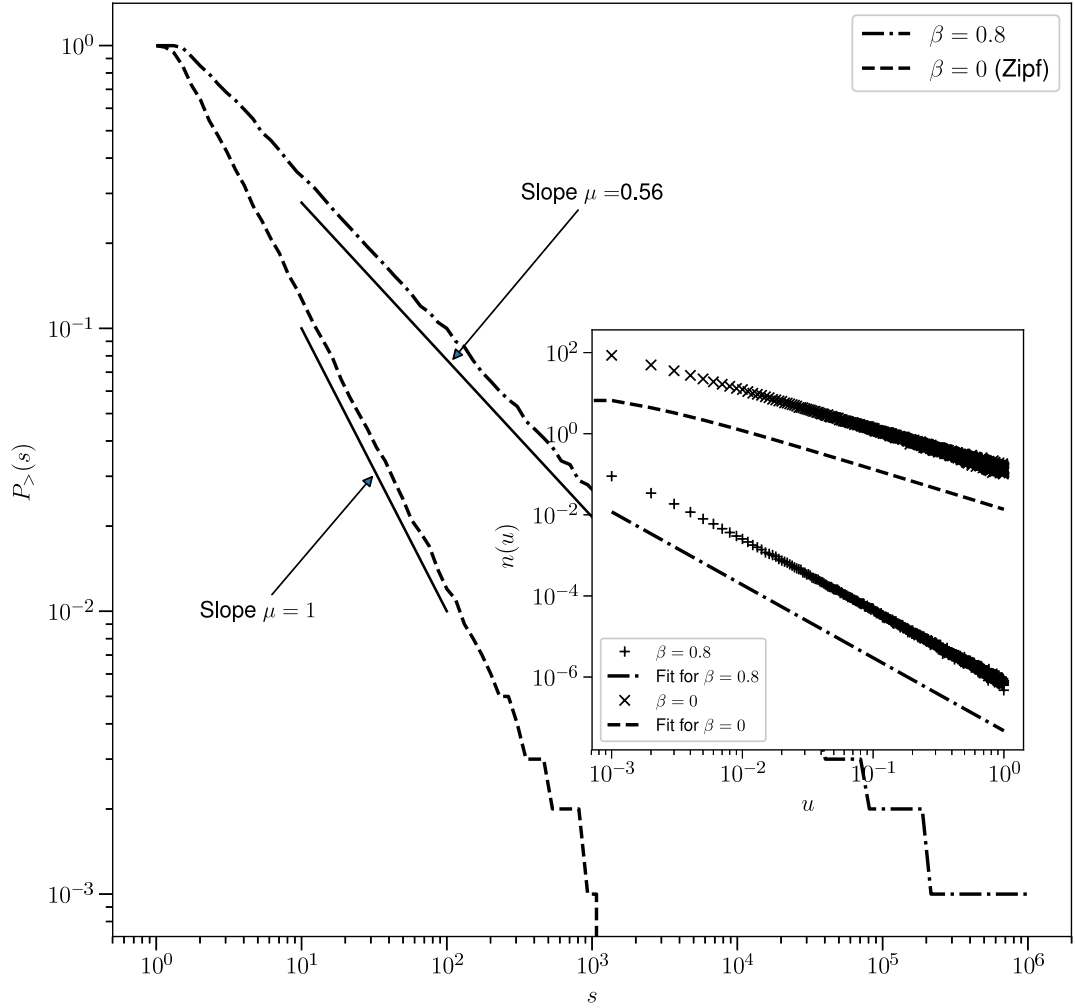}
\caption{\label{fig:ps_nu} Log-log plots of $P_>(s):=\int_s^\infty\dint s'P(s')\sim s^{-\mu}$, with $\mu:=\frac{1}{\beta+1}$ and $n_{\text{st}}(u)$ as given by numerical simulations along with their corresponding fits to the analytical results.}
\end{figure}
 
\section{The multi-criterion case} 
\label{sec:moving_onto_the_multi_dimensional_case}

Let us formally define the model by specifying the rules according to which an agent in a firm characterized by a vector of scores $\vec{x}=(x_1,\ldots,x_K)$ can move into a new firm with scores $\vec{y}=(y_1,\ldots,y_K)$. 
Setting again $u_i=1-x_i$ and $v_i=1-y_i$, the transition is allowed when there exists a subset $(i_1,\ldots,i_M)\in \{1,\ldots,K\}$ such that:
\begin{equation}\label{eq:win_condition}
   \begin{split}
   v_{i_1}&<u_{i_1}\\
   &\vdots\\
   v_{i_M}&<u_{i_M}.
   \end{split}
 \end{equation} 
 If this is the case, the corresponding transition matrix $W_{\vec{u}\rightarrow \vec{v}}$ is set to unity, otherwise $W_{\vec{u}\rightarrow \vec{v}}=0$. 
 
Defining $n(\vec{u},t)$ as the density of agents in firms of score $\vec{u}$ at time $t$, the general Master equation we will consider is
\be\label{eq:masterK}
\frac{\partial n(\vec u,t)}{\partial t} = - \Gamma n(\vec u,t) \omega_{K,M}^\beta(\vec{u}) + \Gamma \int_{ W_{\vec{v}\rightarrow \vec{u}}\neq 0} {\rm d}\vec v \, n(\vec v,t) \omega_{K,M}^{\beta-1}(\vec{u}) - \varphi n(\vec u,t) + \varphi,
\ee
where 
\be
\omega_{K,M}(\vec{u})= \int_{[0,1]^K}\dint \vec{v}\, W_{\vec{u}\rightarrow\vec{v}}
\ee
is the total volume accessible for $\vec{u}$.  The case $\beta=0$ corresponds to the ``greedy'' algorithm of the previous section, while $\beta=1$ corresponds to its ``myopic'' counterpart. 

In the stationary state, one can map models with different values of $\beta$ in the limit $\varphi\to 0$ by writing:

\begin{equation}
  n_{\text{st},\beta\neq0}(\vec{u})\propto \frac{n_{\text{st},\beta=0}(\vec{u})}{\omega_{K,M}(\vec{u})^\beta}.
\end{equation}

\subsection{Grand Slam}

Let us first consider the case $M=K$, i.e. when all scores must improve for the agent to change firm. In this case, the available volume $\omega_{K,K}(\vec{u})$ is easy to compute and reads:
\be
\omega_{K,K}(\vec{u}) = \prod_{i=1}^{K} u_i.
\ee
The stationary distribution then takes a factorized form, which for $\beta=0$ reads:
\be\label{eq:GS}
n_{\text{st}}(\vec u) = \prod_{i=1}^{K} \frac{\phi}{1+\phi}\frac{1}{u_i^{\frac{1}{1+\phi}}}.
\ee
This result is expected, because each of the $K$ scores, in this setting, follows an independent one-dimensional greedy process. 

The stationary firm size distribution is given, for $\beta=0$, by 
\be\label{eq:PsK}
P(s) = \int_{[0,1]^K}\dint \vec{u} \, \delta\left(s - \prod_{i=1}^{K} \frac{\phi}{1+\phi}\frac{1}{u_i^{\frac{1}{1+\phi}}}\right).
\ee

This can be computed by first introducing the following function:
\be
F_K(t) = \int_{[0,1]^K}\dint \vec{u} \, \delta\left(t - \prod_{i=1}^{K} u_i \right).
\ee
Integrating over -- say -- $u_K$, one easily derives the following recursion relation:
\be
F_K(t) = \int_t^1 \dint s\frac{F_{K-1}(s)}{s},
\ee
which solves as
\be\label{eq:fk_sol}
F_K(t) = \frac{(\log t^{-1})^{K-1}}{(K-1)!}.
\ee
Injecting this expression into Eq. (\ref{eq:PsK}) we thus find:
\be
P(s)=\int {\rm d}t F_K(t) \delta\left(s - A_K t^{-\frac{1}{1+\phi}}\right),
\ee
with $A_K=\phi^K/(1+\phi)^K$. The final result is a Zipf law with a $K$ dependent logarithmic correction:
\be
P(s \to \infty) \propto \left(\log \frac{s}{A_K}\right)^{K-1} s^{-2-\phi}.
\ee

\subsection{Marginal distributions}

In the following, we consider $M < K$. As we will see, no regularization is needed in this case since agents can always ``escape'' from good firms. Hence, we set $\varphi = 0$ henceforth.
Taking the distribution of the new scores $\vec{v}$ conditioned on the previous scores $\vec{u}$, and integrating over $(v_1,\ldots, v_{K-1})$ allows us to compute the marginal distribution of $v_K$, $P(v_K)$, given $\vec{u}$
  \begin{equation}\label{eq:marginal}
     \begin{split}
     P(v_K\vert \vec{u})&= \frac{\theta(u_K-v_K)\omega_{K-1,M-1}(\vec u_{\bar K})+\theta(v_K-u_K)\omega_{K-1,M}(\vec u_{\bar K})}{u_K\omega_{K-1,M-1}(\vec u_{\bar K})+(1-u_K) \omega_{K-1,M}(\vec u_{\bar K})}\\
     &= \frac{\theta(u_K-v_K)+\nu(\vec u_{\bar K}) \theta(v_K-u_K)}{u_K+ \nu(\vec u_{\bar K})(1-u_K)},
     \end{split}
  \end{equation}
where $\theta(x)=1$ for $x >0$ and zero otherwise, $\vec u_{\bar K}=(u_1, \ldots, u_{K-1})$ and
\be
\nu(\vec u_{\bar K}):=\frac{\omega_{K-1,M}}{\omega_{K-1,M-1}}
\ee
where $\nu=\frac{\omega_{K-1,M}(\vec u_{\bar K})}{\omega_{K-1,M-1}(\vec u_{\bar K})}$. 

Assuming that $\nu$ converges to a constant value $\nu^*$ (which is in fact only true in the limit $K \to \infty$, see below), the stationary marginal distribution of scores $P_{\text{st}}(v)$ must verify
\begin{equation}\label{eq:p_nu_stat}
\begin{split}
P_{\text{st}}(v)&=\int_{v}^{1}\dint u\, \frac{P_{\text{st}}(u)}{u+\nu^* (1-u)}+\nu^*\int_{0}^{v}\dint u\, \frac{P_{\text{st}}(u)}{u+\nu^* (1-u)}
\end{split}
\end{equation}
which has a (normalized) solution:
\begin{equation}\label{eq:p_nu_sol}
\begin{split}
P_{\text{st}}(u)=\frac{\nu^*-1}{\log(\nu^*)}\frac{1}{u+(1-\nu^*)u}
\end{split}
\end{equation} 

We now show that in the large $K$ limit, $\nu^*$ can be self-consistently computed and only depends on the ratio $\alpha = \frac{M}{K}$.

\subsection{The large $K$ limit}
  
We first note that the volume $\omega_{K,M}(\vec{u})$ can be represented exactly using binary spin variables $\sigma_k=\pm 1$, as
 \begin{equation}\label{eq:vol_def}
    \begin{split}
    \omega_{K,M}(\vec{u})&=\int_{[0,1]^K}\dint \vec{v}\, W_{\vec{u}\rightarrow\vec{v}}\\
    &=\sum_{\{\sigma_k=\pm 1\}}\prod_{k=1}^{K}u_k^{\frac{1+\sigma_k}{2}}(1-u_k)^{\frac{1-\sigma_k}{2}}\theta\left(\sum_{k=1}^{K}\sigma_k-\left(2M-K\right)\right).
    \end{split}
 \end{equation}
 Using the Fourier representation of the Dirac delta function, one finds:
\begin{equation}\label{eq:v_1}
 \begin{split}
    \omega_{K,M}(\vec{u})&\propto\int_{2\alpha-1}^1\mathrm{d}\mu\,\int \dint \lambda\, e^{-i \mu \lambda} \sum_{\{\sigma_k=\pm1\}}\exp\left(\sum_{k=1}^{K}\left(h(u_k)+ i\lambda\right) \sigma_k+\frac{\log(u_k(1-u_k))}{2}\right)
 \end{split}
\end{equation}
  where $h(u)=\frac{1}{2}\log\left(\frac{u}{1-u}\right)$. 
  
\begin{figure}
  \includegraphics[width=1.0\textwidth]{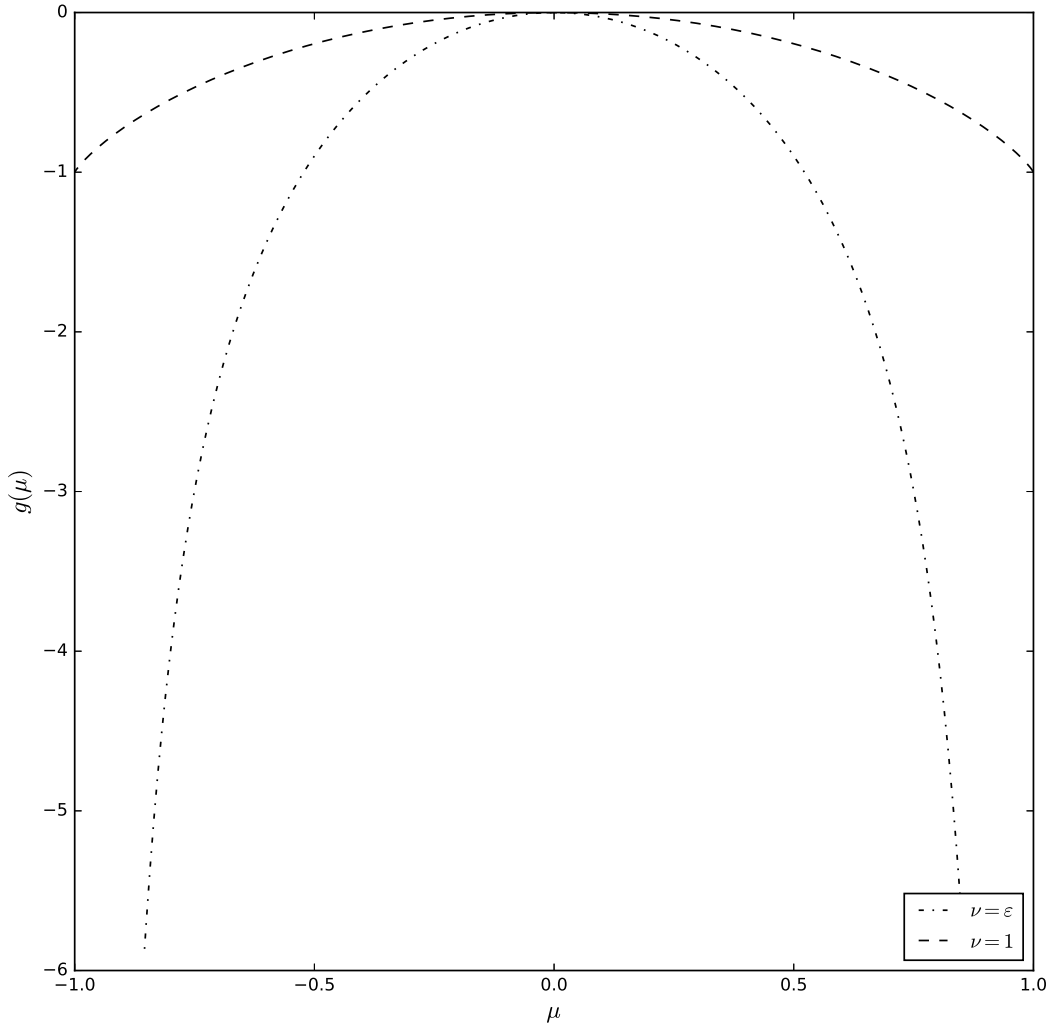}
  \caption{\label{fig:muofg} Plot of $g(\mu)$ as computed for $\nu=1$ and $\nu=\varepsilon$. Note that the maximum is reached in both cases for $g(0)=0$, allowing one to establish the result in Eq. \eqref{eq:closed_epsilon}.}
  \end{figure}

Now we are left with computing the partition function of a system of Ising spins with local fields $h(u_k)+i\lambda$. Summing over all configurations $\{\sigma_k=\pm1\}$ we find that each coordinate contributes to the partition function with a term $\left(\cosh(h(u_k)+i\lambda)\right)$. After a few algebraic manipulations we obtain
  \begin{equation}\label{eq:v_2}
  \begin{split}
  \omega_{K,M}(\vec{u})&\propto \int_{2\alpha-1}^{1}\dint \mu \,\int \dint \lambda \,\exp\left(\sum_{k=1}^K \left(\log\left(u_k+e^{-2i\lambda}(1-u_k)\right)-i\lambda(\mu-1)\right)\right)
  \end{split}
  \end{equation}
In the large $K$ limit two simplifications occur. First, one can replace the sum $\sum_{k=1}^{K} f(u_k)$ with the integral $K\int_{0}^{1}\dint u P(u) f(u)$ when $f$ is an arbitrary regular function. Second, one can calculate the integral over $\lambda$ using a saddle-point method. Hence:
 \begin{equation}\label{eq:1saddle}
  \begin{split}
  \omega_{K,M}(\vec{u})&\propto \int_{2\alpha-1}^{1}\dint \mu \,\exp K\left(\int_{0}^{1}\dint u\,P(u) \left(\log\left(u+e^{-2\lambda^*(\mu)}(1-u)\right)-\lambda^*(\mu)(\mu-1)\right)\right)
  \end{split}
  \end{equation}
  where $\lambda^*(\mu)$ satisfies the saddle-point equation:
  \begin{equation}\label{eq:lambda_saddle}
  \begin{split}
  \mu = \int_{0}^1 \mathrm{d}u P(u) \frac{\tanh(\lambda^*)+2u-1}{1+\tanh(\lambda^*)(2u-1)}. 
  \end{split}
  \end{equation}
 Assuming that $P(u)$ has reached its stationary limit $P_{\text{st}}(u)$ given by Eq. (\ref{eq:p_nu_sol}), we therefore find:
    \begin{equation}\label{eq:lambda_saddle2}
    \mu = \left\{\begin{matrix}
    \frac{2(\nu^*-1)\varepsilon\log(\varepsilon)-(\varepsilon-1)\log(\nu^*)(\nu^*+\varepsilon)}{(\varepsilon-1)\log(\nu^*)(\varepsilon-\nu^*)} &\mbox{ if }& \varepsilon\neq \nu^*\\
    \frac{2}{\log(\nu^*)}-\frac{\nu^*+1}{\nu^*-1} &\mbox{ if }& \varepsilon=\nu^*
    \end{matrix}\right.
  \end{equation}
  with $\varepsilon:=e^{-2\lambda^*}$. This last equation must be solved for $\lambda^*$ for a given pair of values $\mu, \nu^*$. The remaining integral over $\mu$ is again estimated using a second saddle-point, leading to:
  \begin{equation}\label{eq:mumax}
  \begin{split}
  \mu^*(\alpha,\nu^*) &= \underset{\mu\in[2\alpha-1,1]}{\mathrm{argmax}}\left(\int_{0}^{1}\dint u\,P_{\text{st}}(u) \left(\log\left(u+e^{-2\lambda^*(\mu)}(1-u)\right)-\lambda^*(\mu)(\mu-1)\right)\right)\\
  &:=\underset{\mu\in[2\alpha-1,1]}{\mathrm{argmax}}g(\mu,\nu^*)
  \end{split}
  \end{equation}
  and we obtain our final result for $\omega_{K,M}(\vec{u})$:
  \begin{equation}\label{eq:volume}
  \begin{split}
   \omega_{K,M}(\vec{u})&\propto\exp  \left(K\int_{0}^{1}\dint u\,P_{\text{st}}(u) \left(\log\left(u+e^{-2\lambda^*(\mu^*)}(1-u)\right)-\lambda^*(\mu^*)(\mu^*-1)\right)\right)
  \end{split}
  \end{equation}
  One can check that for all $\nu \in [0,1]$, the function $g(\mu)$ is decreasing in the interval $[0,1]$, and has a maximum for $\mu=0$ with $g(0)=0$ as depicted in Fig. \ref{fig:muofg} for $\nu=\varepsilon$ or $\nu=1$. Assuming that the saddle point is $\mu^*=2 \alpha -1$ when $\alpha > \frac12$ and $\mu^*=0$ otherwise, one can compute the relevant volume ratio for large $K$ and for $\alpha > \frac12$ using Eq. \eqref{eq:volume}:
  \be
  \frac{\omega_{K-1,M}}{\omega_{K-1,M}}=\frac{\omega_{K-1,\alpha\left(1+\frac{1}{K}\right)}}{\omega_{K-1,\alpha+\frac{\alpha-1}{K}}} \approx e^{-2\lambda^*(\mu^*)}:=\varepsilon^*.
  \ee
  validating thus our assumption for the saddle-point. Setting $\nu^* = \varepsilon^*$ in Eq. \eqref{eq:lambda_saddle2}, we finally obtain an implicit expression for $\nu^*$:
  \begin{equation}\label{eq:closed_epsilon}
  \begin{split}
  1 - 2 \alpha &= \frac{\nu^*+1}{\nu^*-1} - \frac{2}{\log(\nu^*)} \mbox{ when } \alpha>\frac{1}{2}\\
  \nu^*&=1 \mbox{ when } \alpha\leq\frac{1}{2}.
  \end{split}
  \end{equation}
  
\begin{figure}
  \includegraphics[width=1.0\textwidth]{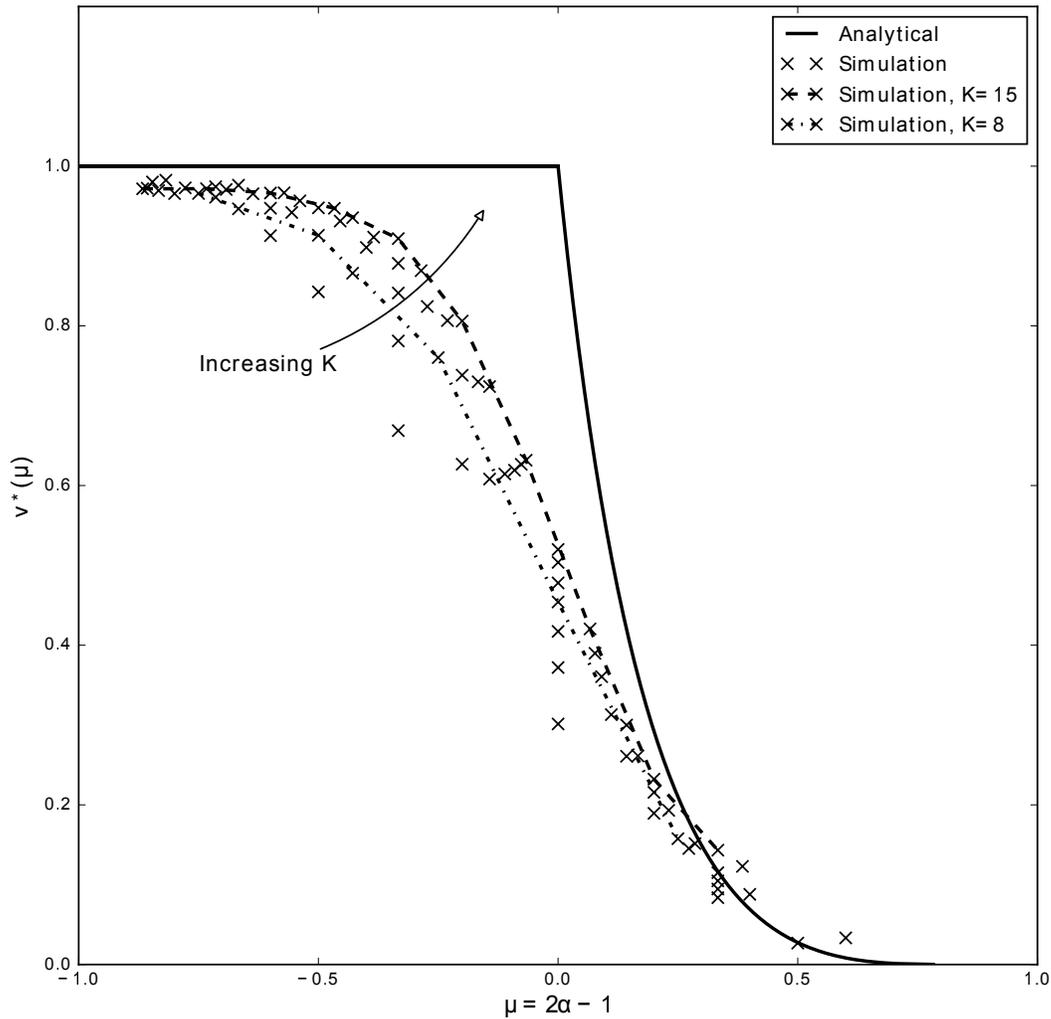}
  \caption{\label{fig:mu_sims} Plot of $\nu^*$ computed using the empirical transition matrix from simulations with $K=2\ldots15$ vs. the analytical value given in Eq. \eqref{eq:closed_epsilon}. Note that the empirical values get closer to the analytical curve as $K$ gets larger, as shown by the dashed lines over the values for $K=15$ and $K=8$.}
  \end{figure}

\begin{figure}
  \centering
  \includegraphics[width=1.0\textwidth]{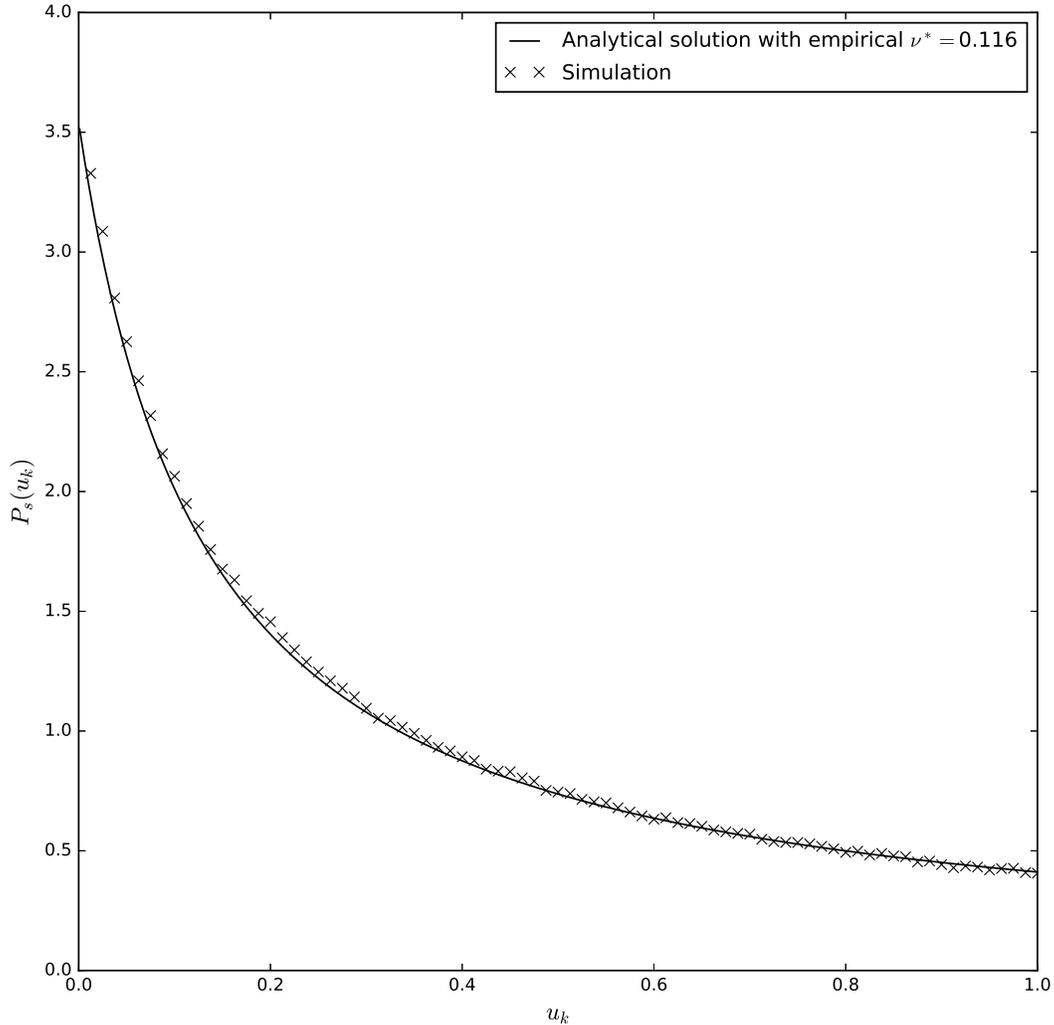}
  \caption{\label{fig:m8k12} Plot of empirical and analytical $P_{\text{st}}(u)$ for $M=8$ and $K=12$. The value of $\nu^*$ was again determined using the empirical transition matrix and plugged into Eq. \eqref{eq:p_nu_sol}.}
  \end{figure}

  We compare our analytical results with numerical simulations in Figs. \ref{fig:mu_sims} and \ref{fig:m8k12}. The shape of $P_{\text{st}}(u)$ describes very well the numerical results even for small values of $K$. The value of $\nu^*$, however, is only slowly reached when $K \to \infty$.

  When $\alpha \to 1$ (i.e. when $M \to K$), the solution for $\nu^*$ is exponentially small:
  \be
  \nu^* \approx e^{-\frac{1}{1-\alpha}},
  \ee
  whereas when $\alpha \to \frac12$ from above,  $\nu^*\approx 1-6(2\alpha-1)$ and tends to $1$. When $\alpha < \frac12$, $\nu^* = 1$ and $P_{\text{st}}(u)=1$. Hence, in the case where $M < K/2$, the distribution of scores is 
  uniform in the large time limit: the selection process is too weak to induce any differentiation between firms. 
  
  Since in the large $K$ limit the dynamics of scores decouple, the full stationary distribution $n_{\text{st}}(\vec u)$ is simply given by 
  \begin{equation}\label{eq:main_res}
  \begin{split}
  n_{\text{st}}(\vec u) \propto \prod_{i=1}^{K}\frac{1}{\left(u_i+(1-u_i)\nu^*\right)^{1+\beta}}
  \end{split}
  \end{equation}
  This is the central result of this section. Notice that the re-injection mechanism that we introduced in the mono-criterion case is not needed here since $\nu^* > 0$ whenever $\alpha < 1$. 
  When $\alpha \to 1$, $\nu^* \to 0$ and the stationary distribution is extremely close to the result for $M=K$, see Eq. (\ref{eq:GS}) for $\phi=0$.

\begin{figure}
\includegraphics[width=1.0\textwidth]{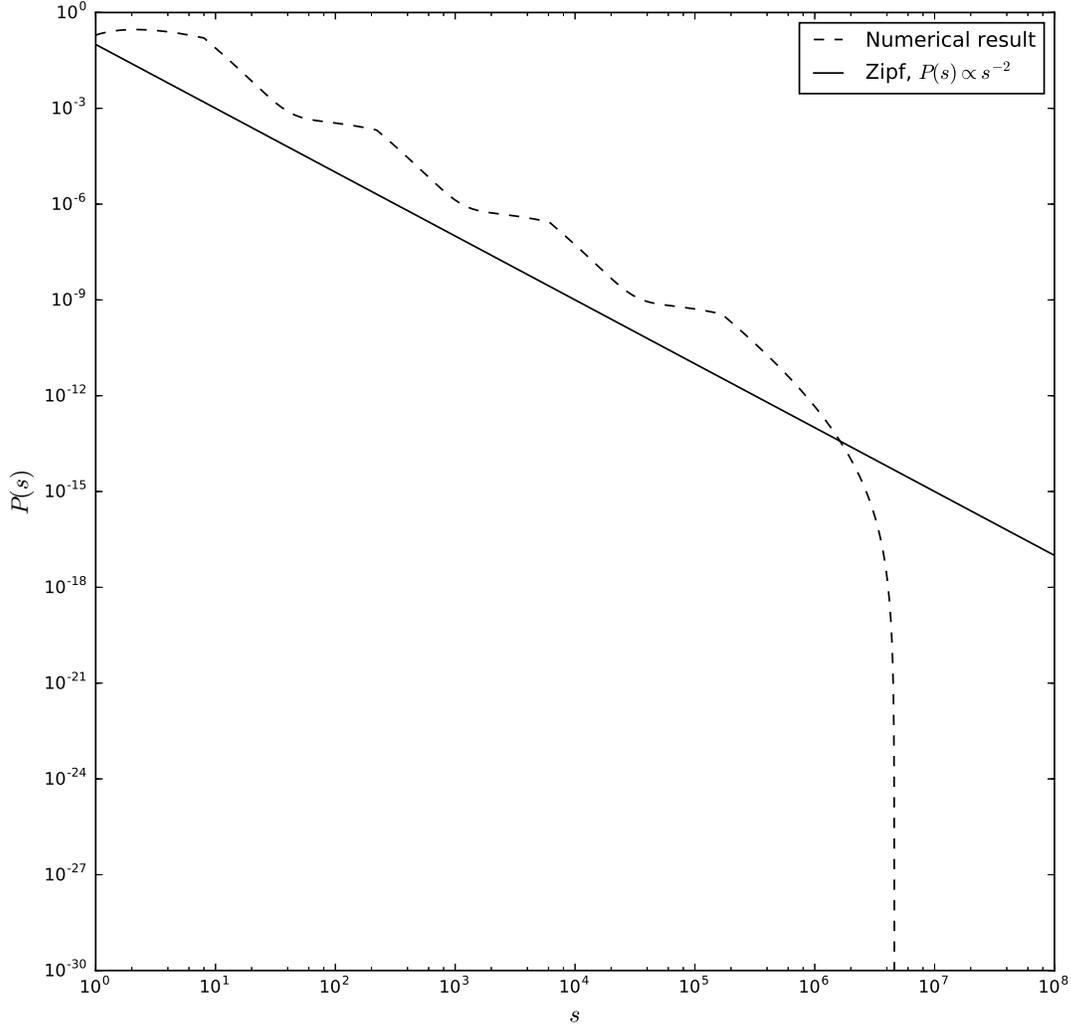}
\caption{\label{fig:ps_example} Log-log plot of $P(s)$ for $K=5$ and $M=4$, compared to the Zipf law $s^{-2}$. The figure was plotted using the numerical value of $\nu^*\simeq 0.036$. Note the modulation introduced by the 
logarithmic corrections encoded in the function $F_K(u,\nu)$.}
\end{figure}

Now the distribution of firm sizes can be computed as
\begin{equation}\label{eq:pn_multidim}
\begin{split}
P(s)&= \int_{[\nu;1]^K}\dint \vec{v}\,\,\delta \left(s-A_K\prod_{i=1}^K v_i\right)\\
&= \int_{[\nu^{K};1]}\dint t \,\delta\left(n-A_Kt^{-1}\right)F_{K}(t,\nu)
\end{split}
\end{equation}
with $v_i=u_i+(1-u_i)\nu^*$, $A_K=\left(\frac{\nu^*-1}{\log(\nu^*)}\right)^K$. We have introduced the function $F_K(t,\nu)=\int_{[\nu;1]^K}\dint \vec{u}\,\delta(t-\prod u_i)$ that satisfies:
\begin{equation}\label{eq:rec_eq}
\left\{
\begin{matrix}
 F_1(t,\nu)&=&\theta(t-\nu)\\
 F_{K}(t,\nu)&=&\int_{t}^{\min\left(\frac{t}{\nu},1\right)}\frac{\dint s}{s}F_{K-1}(s,\nu),
 \end{matrix} \right.
\end{equation}
which can be computed explicitly for small values of $K$:\footnote{Expressed in terms of $\ell_i=\log(u_i+(1-u_i)\nu^*)$, the problem can be mapped onto a sum of independent random variables, uniformly distributed in $[0,1]$. The corresponding distribution for $K > 1$ is called the Irvin-Hall distribution.}
\begin{equation}\label{eq:explicit_fk}
\begin{split}
F_2(t,\nu)=&\theta(t-\nu)\log(t^{-1})+\theta\left(\nu-t\right)\theta\left(t-\nu^2\right)\log\left(\frac{t}{\nu^2}\right)\\
2F_3(t,\nu)=&\theta\left(t-\nu\right) \log^2(t^{-1})+\theta\left(\nu-t\right)\theta\left(t-\nu^2\right)\left(\log\left(\frac{\nu}{t}\right)^2+\log\left(\frac{t}{\nu^2}\right)^2\right)\\
&+\theta\left(\nu^2-t\right)\theta\left(t-\nu^3\right)\log^2\left(\frac{t}{\nu^3}\right).
\end{split}
\end{equation}
One can convince one self that this structure generalises for larger $K$s, with contributions $\frac{\log^{K-1}\left(\frac{\nu^j}{t}\right)}{(K-1)!}$ and $\frac{\log^{K-1}\left(\frac{t}{\nu^{j+1}}\right)}{(K-1)!}$ when  $t\in[\nu^{j+1};\nu^j], \quad j=1\ldots K-1$. In the end intervals one finds $F_K(t,\nu)=\frac{(\log^{K-1}(t^{-1}))}{(K-1)!}$ and $F_K(t)=\frac{\log^{K-1}(\frac{t}{\nu^K})}{(K-1)!}$ for $\nu^{K}<t<\nu^{K-1}$. Finally, 
$F_K(t,\nu)=0$ for $t<\nu^K$. Although quite complicated, one therefore finds that $F_K(t,\nu)$ is a piece-wise polynomial function of $\log t^{-1}$ in the whole interval. Hence, the distribution $P(s)$, given by:
 \begin{equation}\label{eq:final}
   P(s)\propto\left\{
   \begin{matrix}
    s^{-2}F_K(\frac{A_K}{s},\nu) &\mbox{ for }& s< s^* = \frac{A_K}{\nu^{*K}}\\
    0 &\mbox{ for }& s> s^* =\frac{A_K}{\nu^{*K}}, 
    \end{matrix} \right.
 \end{equation}
is a Zipf-law modulated by powers of logarithms in a broad interval that extends up to $s^*\underset{\nu\to 0}{\simeq} \left(\nu\log\left(\frac{1}{\nu}\right)\right)^{-K}$. The important point here is that as soon as $\nu^* < 1$, the upper cut-off $s^*$ becomes very large when $K$ is large, all the more so when $\alpha$ is close to unity. An example of a plot of $P(s)$ is given in Fig. \ref{fig:ps_example}.

\section{Conclusion}

We have introduced a simple model of firm/city/etc. growth based on the idea that agents choose to switch from entity A to entity B depending on a multi-item criterion: whenever entity B fares better that entity A on a subset of $M$ items out of $K$, the agent originally in A moves to B. We have solved the model analytically in the cases $K=1$ and $K \to \infty$. The resulting stationary distribution of sizes is generically a Zipf-law provided $M > K/2$. When $M \leq K/2$, no selection occurs and the size distribution remains thin-tailed. In the special case $M=K$, some regularisation process must be introduced to prevent the whole system from condensating into the ``best'' entity. Introducing a small probability $\phi$ that each entity defaults and redistributes all its agents in surving entitities, one finds that the stationary distribution has a power-law tail that becomes a Zipf-law when $\phi \to 0$. The approach to the stationary state has also been characterized, with strong similarities with a simple ``aging'' model considered by Barrat \& M\'ezard~\cite{barrat1995phase}. Although our model for $K=1$ looks superficially similar to those considered by Gualdi \& Mandel~\cite{gualdi2016emergence} and Axtell~\cite{axtell2013endogenous}, we have not been able to elicit a precise mapping between these models, and more generally with the slew of other mechanisms that lead to Zipf-laws. 


Acknowledgements: We benefited from very fruitful discussions with M. Cordi, S. Gualdi, M. Tarzia and F. Zamponi, who participated actively to the first stages of this project.

  \bibliography{biblio.bib}

\end{document}